\documentclass[english,aps,prd,preprint,amsfonts,amsmath,amssymb,superscriptaddress]{revtex4-1}
\usepackage[T1]{fontenc}
\usepackage[latin9]{inputenc}
\usepackage{color}
\usepackage{verbatim}
\usepackage{amsmath}
\usepackage{amssymb}
\usepackage{graphicx}
\usepackage{esint}
\usepackage{babel}

\begin{document}

\title{Secular Gravity Gradients in Non-Dynamical Chern-Simons Modified Gravity for Satellite Gradiometry Measurements}

\author{Li-E Qiang}
\email{qqllee815@chd.edu.cn}

\affiliation{Department of Geophysics, College of the Geology Engineering and
Geomatics, Chang'an University, Xi'an, 710054, China.}


\begin{abstract}
With continuous advances in related technologies, relativistic gravitational experiments with orbiting gradiometers becomes feasible, which could naturally be incorporated into future satellite gravity missions. Tests of Chern-Simons modified gravity are meaningful since such a modification gives us insights into (possible) parity-violations in gravitation. In this work, we derive, at the post-Newtonian level, the new observables of secular gradients from the non-dynamical Chern-Simons modified gravity, which will greatly improve the constraint on the mass scale $M_{CS}$ that may be drawn from satellite gradiometry measurements. For superconducting gradiometers, a strong bound $M_{CS}\gtrsim 10^{-7}\ eV$ could in principle be obtained. For future optical gradiometers based on similar technologies from the LISA PathFinder mission, a even stronger bound $M_{CS}\gtrsim 10^{-6}\sim10^{-5}\ eV$ might be expected.
\end{abstract}

\maketitle

\section{Introduction}

Among modifications to Einstein's General Relativity (GR), extensions to the Einstein-Hilbert action with second order curvature terms are of particular interest, which may arise from the full, but still lacking, quantum theory of gravity \cite{Niedermaier2006}. The  Chern-Simons (CS) modified gravity \cite{Deser1982,Campbell1990,Campbell1991,Jackiw2003,Alexander2009} is such an extension of GR, which had found connections to string theory, loop quantum gravity, particle physics and etc, see \cite{Alexander2009} for detailed discussions. More interestingly, the CS modified gravity may give us insights into (possible) parity-violations in gravitation, therefore, its experimental tests and the resulted constraints are of importance. The first constraint \cite{Smith2008} on the mass scale $M_{CS}$ of the non-dynamical formulation of CS modified gravity was obtained based on the observations from the LAGEOS I, II \cite{Ciufolini2004,Ciufolini2007} and the Gravity Probe-B \cite{Everitt2011} missions, which had set $M_{CS} \gtrsim 10^{-13}\ eV$, and up to now the strongest bounds $M_{CS}\gtrsim 4.7\times10^{-10}\ eV$ (been revised in \cite{Ali-Haimoud2011a}) was based on the data from double binary pulsars \cite{Yunes2009a}.

In recent works \cite{Qiang2015,Qiang2015a}, the authors had suggested a new method of testing CS modified gravity with orbiting gradiometers, which could be incorporated into future satellite gravity missions. With the success of the LISA PathFinder mission \cite{Armano2016}, which can be view as a demonstration of an one dimensional optical gradiometer with the resolution of the laser interferometer better than $9\ pm/\sqrt{Hz}$ in the $mHz$ band and the noise floor $4\sim5$ orders of magnitude below the best level achieved by GOCE \cite{Rummel2011}, and the continuous advances in superconducting gradiometers \cite{Moody2002,Griggs2015}, precision tests of relativistic gravitational theories including the CS modified gravity with satellite gradiometry becomes more and more feasible. Relativistic gravitational experiments with orbiting gradiometers was first studied in 1980s \cite{Mashhoon1982,Mashhoon1989,Paik1989}, and it is noticed by Mashhoon and Theiss \cite{Mashhoon1982,Mashhoon1984,Mashhoon1985,Theiss1985} that the existence of secular gravity gradients or tidal effects in local free-falling frames along orbit motions (known as the Mashhoon-Theiss anomaly) would greatly improve the measurement accuracy. Recently, the physical mechanism behind such secular tidal effects had been clarified in \cite{Xu2016}, that the relativistic precessions of the local free-falling frame and the orbit plane with respect to the sidereal frame will produce modulations of Newtonian tidal forces and then gives rise to secular tidal forces along certain axes \cite{Xu2016,Qiang2016}. Back to the tests of the CS modified gravity, in the far field expansion of the non-dynamical theory, the CS gravity will add modifications to the GravitoMagnetic (GM) sector of GR \cite{Alexander2007,Alexander2007a}, and therefore will give rise to new secular tidal effects that could be read out precisely along certain axes of an orbiting gradiometer. In this letter, we derive, at the Post-Newtonian (PN) level, the new secular tidal tensor from the non-dynamical CS modified gravity under the local Earth pointing frame along a relativistic polar and nearly circular orbit. For (possible) future experiments, we give the estimations on the bound of $M_{CS}$  that could be drawn from such a measurement scheme.

\section{Models and Settings}
The action of the CS modified gravity is given by
\begin{equation}
S =\frac{1}{16\pi}\int d^{4}x\sqrt{-g}(R+4\pi\alpha\theta R^{\star}R+\mathcal{L}_{matter}),
\end{equation}
for clarity the units $G=c=\hbar=1$ are adopted hereafter and in the end the SI units will be recovered. In the non-dynamical theory with the canonical coupling, the scalar $\theta$ is an externally prescribed and spatially isotropic function that proportional to the coordinate time. The field equation reads
\begin{equation}
R_{\mu\nu}-\frac{1}{2}g_{\mu\nu}R+16\pi\alpha C_{\mu\nu}=8\pi T_{\mu\nu},\label{eq:field_eq}
\end{equation}
where
\begin{equation}
C^{\mu\nu}=\nabla_{\rho}\theta\epsilon^{\rho\lambda\sigma(\mu}\nabla_{\sigma}R_{\ \; \lambda}^{\nu)}+\frac{1}{2}\nabla_{\rho}\nabla_{\lambda}\theta\epsilon^{\sigma \delta\lambda(\mu}R_{\ \ \ \delta\sigma}^{\nu)\rho}.\nonumber
\end{equation}
The introduction of the new scalar degree of freedom gives rise to a new constraint
\begin{equation}
\nabla_{\nu}C^{\nu\mu}=-\frac{1}{8}\nabla^{\mu}\theta(^{\star}RR)=0.\nonumber
\end{equation}
If the constraint is satisfied, the Bianchi identities and the equations of motion for matter fields $\nabla_{\nu}T^{\nu\mu}=0$ are recovered, which ranks the non-dynamical CS modified gravity a metric theory \cite{Will2014}.

In this letter, we model Earth as an ideal uniform and rotating spherical body with total mass $M$ and angular momentum $\vec{J}$. The geocentric inertial coordinates system $\{t,\ x^i\}$ is defined as follows, that one of its bases $\frac{\partial}{\partial x^3}$ is parallel to the direction of $\vec{J}$ and the coordinate time $t$ is measured in asymptotically flat regions. For an orbiting proof mass or satellite, we have the PN order relations
\begin{equation}
v^2\sim\frac{M}{r}\sim\mathcal{O}(\epsilon^2),\ \ \ \  \frac{Jv}{r^2}\sim \mathcal{O}(\epsilon^4),\label{eq:PNorder}
\end{equation}
where $\vec{v}$ is the 3-velocity, $r=\sqrt{\sum_{i=1}^3(x^i)^2}$ and for low (with altitude below $2000\ km$) and medium (altitude between $2000\ km$ and a geostationary orbit) Earth orbits $\epsilon=\frac{M}{r}$ is about $10^{-5}\sim 10^{-6}$. Up to the required order, the metric field outside the ideal Earth model can be expanded as
\begin{equation}
 g_{\mu\nu}=
 \left(\begin{array}{cccc}
-1+\frac{2M}{r}-\frac{2M^{2}}{r^{2}} & (\frac{2 x^{2}}{r^{3}}+\frac{6 x^{1}x^{3}}{r^{5}M_{CS}})J & (-\frac{2 x^{1}}{r^{3}}+\frac{6 x^{2}x^{3}}{r^{5}M_{CS}})J & -\frac{2[(x^{1})^{2}+(x^{2})^{2}-2(x^{3})^{2}]}{r^{5}M_{CS}}J\\
(\frac{2 x^{2}}{r^{3}}+\frac{6 x^{1}x^{3}}{r^{5}M_{CS}})J & 1+\frac{2 M}{r} & 0 & 0\\
(-\frac{ x^{1}}{r^{3}}+\frac{6 x^{2}x^{3}}{r^{5}M_{CS}})J & 0 & 1+\frac{2 M}{r} & 0\\
-\frac{2[(x^{1})^{2}+(x^{2})^{2}-2(x^{3})^{2}]}{r^{5}M_{CS}}J & 0 & 0 & 1+\frac{2 M}{r}
\end{array}\right),\label{eq:metric}
\end{equation}
where the mass scale $M_{CS}$ of the CS modified gravity reads
\begin{equation}
M_{CS}\equiv\frac{1}{8\pi\alpha\dot{\theta}}. \label{eq:mass_scale}
\end{equation}
As mentioned, the non-dynamical CS modified gravity differs from GR only in the GM sector.

\section{Reference orbit and local tetrad}
Being a metric theory, motions of free-falling masses or satellites in CS modified gravity satisfy the geodesic equation.  According to the general choices of orbits for satellite gradiometry missions (like in GOCE \cite{Rummel2011}), we choose the reference orbit followed by the mass center of the gradiometer to be a polar and circular one with the relativistic precession caused by the GM effect
\begin{eqnarray}
x^1&=&a \cos \Psi \cos \dot{\Omega}\tau,\label{eq:x1}\\
x^2&=&a \cos \Psi \sin\dot{\Omega}\tau,\label{eq:x2}\\
x^2&=&a\sin \Psi.\label{eq:x3}
\end{eqnarray}
Here $a$ denotes the orbit radius, $\Psi =\omega\tau$ the true anomaly, $\omega$ is the mean angular frequency with respect to the proper time $\tau$ along the orbit and $\Omega$ is the longitude of ascending note with initial value $\Omega(0)=0$.  The precession rate of the note $\dot{\Omega}=\dot{\Omega}_{GR}+\dot{\Omega}_{CS}$, where the Lense-Thirring precession rate $\dot{\Omega}_{GR}=\frac{2J}{a^3}$ \cite{Lense1918} and the correction from the non-dynamical CS modified gravity had been worked out in \cite{Smith2008} as
\begin{equation}
\dot{\Omega}_{CS}=\Pi_{CS}\dot{\Omega}_{GR}=15\frac{a^2}{R^2}j_2(\frac{R}{a}\frac{4}{\chi})y_1(\frac{4}{\chi})\dot{\Omega}_{GR},\label{eq:OCS}
\end{equation}
where $R$ is the averaged radius of Earth, $j_l(x)$ and $y_l(x)$ are spherical Bessel functions of the first and second kind, and $\chi$ was PN parameter introduced in \cite{Alexander2007,Alexander2007a} which can be related to the mass scale as $\chi=\frac{4}{a M_{CS}}$. For polar circular orbit, the GM force in CS modified gravity generated by spherical sources \cite{Smith2008} will also change slightly the orbital eccentricity to $e\sim \chi\mathcal{O}(\epsilon^2)$, which is too small to be relevant to secular tidal effects at the PN level. Therefore, in this letter the small eccentricity is ignored, and its effects together with other orbital perturbations, such as those from geopotential multipoles, are left for future studies.

For satellite gradiometry missions, spacecraft attitudes are generally chosen to follow the Earth pointing orientation (like GOCE \cite{Rummel2011}). Then, we define the local free-falling Earth pointing frame by the tetrad $\{E_{(a)}^{\ \ \ \mu}\}$ attached to the mass center of the orbiting gradiometer. We set $E_{(0)}^{\ \ \ \mu}=\tau^{\mu}$ with $\tau^{\mu}$ the 4-velocity of the mass center, $E_{(1)}^{\ \ \ \mu}$ is along the 3-velocity $\vec{v}$ in space, $E_{(2)}^{\ \ \ \mu}$ along the radial direction and $E_{(3)}^{\ \ \ \mu}$ is transverse to the orbit plane. For the existence of the geodetic and frame-dragging precession of the local frame \cite{Schiff1960}, we solve for the spatial bases $\{E_{(i)}^{\ \ \ \mu}\}$ in the following three steps. First, in the geocentric coordinates system, we solve for the precession of the local inertial frame (Fermi-shifted frame) along the orbit given in eq.~(\ref{eq:x2})-(\ref{eq:x3}). Second, with respect to the local inertial frame, we rotate $\{E_{(i)}^{\ \ \ \mu}\}$ with an initial angular velocity to make it an Earth pointing triad. At last, since the local frame is moving along the orbit, we need to perform the boost Lorentz transformations of the bases $\{E_{(i)}^{\ \ \ \mu}\}$ with respect to the 4-velocity $\tau^{\mu}$. The general time scales or periods of frame-dragging precessions in Earth orbit are about $\frac{c^2a^3}{GJ}\sim10^7\ yrs$,  which is extremely long compared with general mission lifetimes.  Then, following the above three steps and within the short time limit $\frac{\tau}{a}\ll \frac{a^2}{J}$, the tetrad can be worked out up to the PN level as
\begin{equation}
E_{(0)}^{\ \ \ \mu}=\left(\begin{array}{c}
1+\frac{a^2\omega^2}{2}+\frac{M}{a}\\
-a\omega\sin\Psi \\
0\\
a\omega\cos\Psi
\end{array}\right).\label{eq:E0}
\end{equation}
\begin{equation}
E_{(1)}^{\ \ \ \mu}=\left(\begin{array}{c}
a\omega\\
-(1+\frac{a^2\omega^2}{a}-\frac{M}{a})\sin\Psi\\
-\frac{(1+\Delta_{CS})J\Psi\sin\Psi}{2a^3\omega}\\
(1+\frac{a^2\omega^2}{a}-\frac{M}{a})\cos\Psi
\end{array}\right),\label{eq:E1}
\end{equation}
\begin{equation}
E_{(2)}^{\ \ \ \mu}=\left(\begin{array}{c}
0\\
(1-\frac{M}{a})\cos \Psi\\
\frac{(1+\Delta_{CS})J(\Psi\cos\Psi-3\sin\Psi)}{2a^3\omega}\\
(1-\frac{M}{a})\sin\Psi
\end{array}\right),\label{eq:E2}
\end{equation}
\begin{equation}
E_{(3)}^{\ \ \ \mu}=\left(\begin{array}{c}
0\\
\frac{(1+\Delta_{CS})J(3\sin2\Psi-2\Psi)}{4a^3\omega}\\
1-\frac{M}{a}\\
\frac{3(1+\Delta_{CS})J\sin^2\Psi}{2a^3\omega}
\end{array}\right),\label{eq:E3}
\end{equation}
where the correction to the precessions of the bases from the non-dynamical CS modified gravity reads \cite{Smith2008}
\begin{equation}
\Delta_{CS}=15\frac{a^2}{R^2}j_2(\frac{R}{a}\frac{4}{\chi})[y_1(\frac{4}{\chi})+\frac{4}{\chi}y_0(\frac{4}{\chi})].\label{eq:phiCS}
\end{equation}
\section{Secular gradient observables}
We introduce the position difference vector $Z^{\mu}$ between two adjacent free-falling proof masses in the orbiting gradiometer. Generally $|Z|\sim10^{-1}\ m$, which is much shorter compared with the orbital radius $a\sim10^{7}\ m$, therefore the relative motion between the test masses can be obtained by integrating the geodesic deviation equation along the reference orbit
\begin{equation}
\tau^{\rho}\nabla_{\rho}\tau^{\lambda}\nabla_{\lambda}Z^{\mu}+R_{\rho\nu\lambda}^{\ \ \ \  \mu}\tau^{\rho}\tau^{\lambda}Z^{\nu}=0.\label{eq:deviation}
\end{equation}
In the local frame $\{E_{(a)}^{\ \ \ \mu}\}$, the above geodesic deviation equation can be expanded as
\begin{eqnarray}
\frac{d^{2}}{d\tau^{2}}Z^{(a)}
 & = & -2\gamma_{\ \ \  (b)(0)}^{(a)}\frac{d}{d\tau}Z^{(b)}\nonumber\\
 &&-(\frac{d}{d\tau}\gamma_{\ \ \ (b)(0)}^{(a)}+\gamma_{\ \ \ (b)(0)}^{(c)}\gamma_{\ \ \ (c)(0)}^{(a)})Z^{(b)}\nonumber \\
 &  & -K_{(b)}^{\ \  (a)}Z^{(b)}.\label{eq:localdev}
\end{eqnarray}
where $Z^{(a)}E_{(a)}^{\ \ \ \mu}=Z^{\mu}$ $\gamma_{\ \ \ (b)(c)}^{(a)}=E^{(a)\nu}\nabla_{\mu}E_{(b)\nu}E_{(c)}^{\ \ \ \mu}$ are the Ricci rotation coefficients \cite{Chandrasekhar1983}. The first line of the right hand side of the above equation is the relativistic analogue of the Coriolis force, the second line contains the inertial tidal forces and the last line is the tidal force from the spacetime curvature, where the tidal matrix from curvature is defined by
\begin{eqnarray}
K_{\nu}^{\ \mu} & = & R_{\rho\nu\lambda}^{\ \ \ \ \mu}\tau^{\rho}\tau^{\lambda}.\label{eq:Kdefinition}
\end{eqnarray}
For electrostatic and superconducting gradiometers, the motions of
test masses are suppressed by compensating forces. Then the total
tidal tensor $T_{(a)(b)}$ affecting the gradiometer will be
\begin{equation}
T_{(a)(b)}=-\frac{d}{d\tau}\gamma_{(a)(b)(0)}-\gamma_{(a)(c)(0)}\gamma_{\ \ \ (b)(0)}^{(c)}-K_{(a)(b)}.\label{eq:Tab}
\end{equation}

After straightforward but tedious algebraic manipulations and leaving out all the terms beyond $\frac{1}{a^{2}}\mathcal{O}(\epsilon^{4})$ and $\frac{1}{a^{2}}\Psi\mathcal{O}(\epsilon^{4})$, we work out, to the PN level, the tidal tensors in the local free-falling Earth pointing frame along the reference orbit. For the tidal tensor $K_{(a)(b)}$, as expected we have  $K_{(a)(0)}=0$, and the Newtonian part
\begin{equation}
K_{(i)(j)}^N=\left(
\begin{array}{ccc}
 \frac{M}{a^3} & 0 & 0 \\
 0 & -\frac{2 M}{a^3} & 0 \\
 0 & 0 & \frac{M}{a^3} \\
\end{array}
\right)\label{eq:KN}
\end{equation}
which agrees exactly with the classical Newtonian tidal tensor $\partial_i\partial_j \frac{M}{r}$ evaluated in such local frame. The PN part may be divided into the tidal tensor $K_{(i)(j)}^{GR}$ from GR and the new tensor $K_{(i)(j)}^{CS}$ from the CS modification, which, within the short time limit $\frac{\tau}{a}\ll \frac{a^2}{J}$, can be worked out as
\begin{equation}
K_{(i)(j)}^{GR}=\left(
\begin{array}{ccc}
 -\frac{3 M^2}{a^4} & 0 & \frac{3 J \omega  \cos\Psi}{a^3} \\
 0 & -\frac{3 M \left(a^3 \omega ^2-2 M\right)}{a^4} & -\frac{9 J \left(M \Psi  \cos \Psi+\left(2 \omega ^2 a^3+M\right) \sin \Psi\right)}{2 a^6 \omega } \\
 \frac{3 J \omega  \cos \Psi}{a^3} & -\frac{9 J \left(M \Psi  \cos \Psi+\left(2 \omega ^2 a^3+M\right) \sin \Psi\right)}{2 a^6 \omega } & \frac{3 M \left(a^3 \omega ^2-M\right)}{a^4} \\
\end{array}
\right),\label{eq:KGR}
\end{equation}
\begin{equation}
K_{(i)(j)}^{CS}=\left(
\begin{array}{ccc}
 0 & 0 & 0 \\
 0 &0 & \frac{3 J M (\Psi  (\Delta_{CS} -4 \Pi_{CS}) \cos \Psi-3 \Delta_{CS} \sin \Psi )}{2 a^6 \omega } \\
 0 & \frac{3 J M (\Psi  (\Delta_{CS}-4 \Pi_{CS}) \cos \Psi-3 \Delta_{CS} \sin \Psi )}{2 a^6 \omega } & 0 \\
\end{array}
\right),\label{eq:KCS}
\end{equation}
here $K_{(i)(j)}^{GR}$ agrees exactly with the former result derived in \cite{Xu2016}.
Due to the relativistic precessions of the free-falling local frame and the orbit plane, the modulations of Newtonian tidal tensor given in eq.~(\ref{eq:KN}) produces secular terms in the $K_{(2)(3)}$ and $K_{(3)(2)}$ components, which are the expected secular gradient observables appeared along polar and nearly circular orbits. Finally, with the tidal tensor from inertial forces in eq.~(\ref{eq:Tab}) been worked out,
the total tidal tensor $T_{(i)(j)}$ turns out to be
\begin{equation}
T_{(i)(j)}=\left(\begin{array}{ccc}
-a^{2}\omega^{4}+\frac{4M\omega^{2}}{a}-\omega^{2} & 0& -\frac{J(\Delta_{CS}-3)\omega\cos\Psi}{a^{3}}\\
+\frac{M}{a^{3}}-\frac{3M^{2}}{a^{4}}\\
\\
0 & \frac{7M^{2}}{a^{4}}-\frac{\omega^{2}M}{a}-\frac{2M}{a^{3}}-\omega^{2} & \frac{3JM\Psi(\Delta_{CS}-4\Pi_{CS}-3)\cos\Psi}{2a^{6}\omega}\\
 &  & -\frac{J\left(2(\Delta_{CS}+9)\omega^{2}a^{3}+9M(\Delta_{CS}+1)\right)\sin\Psi}{2a^{6}\omega}\\
\\
\frac{3J(\Delta_{CS}+1)\omega\cos\Psi}{a^{3}} & \frac{3JM\Psi(\Delta_{CS}-4\Pi_{CS}-3)\cos\Psi}{2a^{6}\omega} & -\frac{3M^{2}}{a^{4}}+\frac{3\omega^{2}M}{a}+\frac{M}{a^{3}}\\
 & -\frac{J\left(\left(2(3\Delta_{CS}+9)\omega^{2}a^{3}+9M(\Delta_{CS}+1)\right)\sin\Psi\right)}{2a^{6}\omega} &
\end{array}\right).\label{eq:T}
\end{equation}

\section{Conclusions}

\begin{figure}
\centering
\includegraphics[scale=0.6]{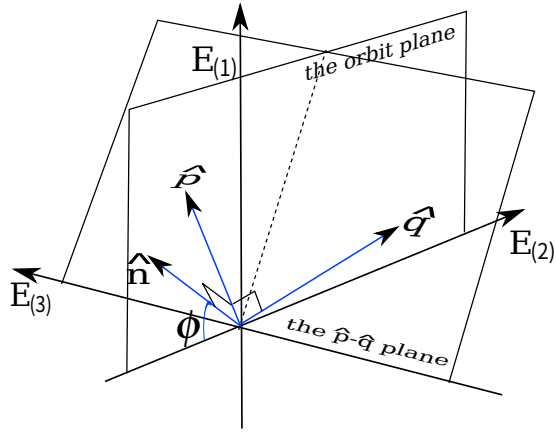}
\caption{The axes of the gradiometer are oriented as follows, $\hat{\mathbf{p}}$
and $\hat{\mathbf{q}}$ are symmetric with respect to the $E_{(1)}^{\ \ \
\mu}-E_{(2)}^{\ \ \ \mu}$ plane, and $\hat{\mathbf{n}}$ is orthogonal to the
$\hat{\mathbf{p}}-\hat{\mathbf{q}}$ plane. The angle between  $\hat{\mathbf{n}}$
and $-E_{(2)}^{\ \ \ \mu}$ is $\phi$.}
\label{fig:3-axis}
\end{figure}

In conclusion, how the new secular gradients given in eq.~(\ref{eq:KCS}) can be read out by an orbiting 3-axis gradiometer is discussed. Following \cite{Mashhoon1989,Paik2008,Qiang2015}, we orient two of the three gradiometer axes 45 degrees above and below the orbital plane and difference their outputs to reject the Newtonian and PN gravitoelectric terms and therefore measure only the GM and secular terms. In the local frame $\{E_{(a)}^{\ \ \ \mu}\}$, the three axes of the gradiometer are oriented as
\begin{eqnarray}
\hat{\mathbf{n}} =  \left(\begin{array}{c}
\sin\phi\\
-\cos\phi\\
0
\end{array}\right),\
\hat{\mathbf{p}}  =  \frac{1}{\sqrt{2}}\left(\begin{array}{c}
\cos\phi\\
\sin\phi\\
-1
\end{array}\right),\
\hat{\mathbf{q}} =  \frac{1}{\sqrt{2}}\left(\begin{array}{c}
\cos\phi\\
\sin\phi\\
1
\end{array}\right),\nonumber
\end{eqnarray}
see Fig.~(\ref{fig:3-axis}) for the illustration.
The difference between the readouts in the $\hat{\mathbf{p}}$ and $\hat{\mathbf{q}}$ axes turns out to be
\begin{eqnarray}
&&\frac{1}{2}(T_{\hat{\mathbf{p}}\hat{\mathbf{p}}}-T_{\hat{\mathbf{q}}\hat{\mathbf{q}}})\nonumber\\
&=&\boxed{-\frac{3J\sin\phi M\Psi(\Delta_{CS}-4\Pi_{CS})\cos\Psi}{2a^{6}\omega}} +\frac{9J\sin\phi M\Psi\cos\Psi}{2a^{6}\omega}\nonumber\\
&&+\frac{J\sin\phi\sin\Psi\left(2a^{3}(2\Delta_{CS}+9)\omega^{2}+9M(\Delta_{CS}+1)\right)}{2a^{6}\omega}\nonumber\\
&&-\frac{J(\Delta_{CS}+3)\omega\cos\phi\cos\Psi}{a^{3}}.\label{eq:Tpq}
\end{eqnarray}
The boxed term is the secular gradient signal $s^{CS}$ from the non-dynamical CS modifications, which grows linearly with time. Errors in such combination $\frac{1}{2}(T_{\hat{\mathbf{p}}\hat{\mathbf{p}}}-T_{\hat{\mathbf{q}}\hat{\mathbf{q}}})$ may arise from misalignments and mispointings of the gradiometer axes, and the related analysis and possible solutions are discussed in \cite{Paik2008}.

Recovering the SI units, we have
\begin{equation}
  s^{CS}=-\frac{3G^2J M\Psi(\Delta_{CS}-4\Pi_{CS})\sin\phi\cos\Psi}{2c^2a^{6}\omega}.\nonumber
\end{equation}
Form previous experiments \cite{Smith2008,Yunes2009a,Ali-Haimoud2011a}, the PN parameter $\chi=\frac{4\hbar c}{a M_{CS}}$ had already been constrained to be a small quantity $\chi\lesssim 10^{-4}$. $\Pi_{CS}$ and $\Delta_{CS}$ can then be expanded as
\begin{eqnarray}
  \Pi_{CS}&=&\mathcal{O}(\chi^2),\nonumber\\
  \Delta_{CS}&=&\frac{15 a^3 \chi  \cos \left(\frac{4}{\chi }\right) \sin \left(\frac{4 R}{a \chi }\right)}{4 R^3}+\mathcal{O}(\chi^2).\nonumber
\end{eqnarray}
Therefore, one has
\begin{eqnarray}
  s^{CS}
=-\frac{45G^2\cos \left(\frac{4}{\chi }\right)\sin \left(\frac{4 R}{a \chi }\right)\chi J M\Psi\sin\phi\cos\Psi}{8c^2a^{3}R^3\omega}+\mathcal{O}(\chi^2).\nonumber
\end{eqnarray}
To give the estimation, we assume the orbital altitude to be $500\ km$ and the mission life time about one year. After one year's accumulation, the total orbital cycle $\Psi$ in $s^{CS}$ will be $3.5\times10^4$, and the secular signal will reach about $4.3\chi\ mE$. With proper data analysis methods employed (such as the matched filtering), the total signal-to-noise ratio can be further amplified by a factor of the square root of the total cycles $\sqrt{\Psi}$. Therefore,
for superconducting gradiometers with potential sensitivity better than $10^{-2}\ mE/\sqrt{Hz}$ in low frequency band near $0.1\ mHz$ \cite{Griggs2015}, a rather strong constraint on the CS mass scale of the non-dynamical theory may in principle be obtained as
\begin{equation}
M_{CS}\gtrsim 10^{-7}\ eV.\nonumber
\end{equation}
For future optical gradiometers based on similar measurement schemes and techniques from the LPF mission, an even stronger bound may be expected
\begin{equation}
  M_{CS} \gtrsim 10^{-6}\sim10^{-5}\ eV.\nonumber
\end{equation}

\begin{acknowledgements}
Supports from National Space Science Center, Chinese
Academy of Sciences (XDA04077700), National Natural
Science Foundation of China (No. 11305255)
and  Central Universities Funds (No. 310826161010) are acknowledged.
\end{acknowledgements}


\end{document}